\documentclass[fleqn]{2017SCGE}
\usepackage{multirow}
\usepackage{amsmath}

\begin{document}

\ensubject{subject}

\ArticleType{Article}
\Year{2021}
\Month{January}
\Vol{60}
\No{1}
\DOI{10.1007/s11432-016-0037-0}
\ArtNo{000000}
\ReceiveDate{January 11, 2021}
\AcceptDate{April 6, 2021}

\title{Modeling the arc and ring structures in the HD 143006 disk}{Modeling the arc and ring structures in the HD 143006 disk}

\author[1]{Yao Liu}{{yliu@pmo.ac.cn}}
\author[2]{Mario Flock}{}
\author[1]{Min Fang}{}

\AuthorMark{Liu, Y.}

\AuthorCitation{Liu, Y.; Flock, M; Fang, M.}

\address[1]{Purple Mountain Observatory \& Key Laboratory for Radio Astronomy, Chinese Academy of Sciences, Nanjing 210023, China}
\address[2]{Max-Planck-Institut f\"ur Astronomie, K\"onigstuhl 17, D-69117 Heidelberg, Germany}


\abstract{Rings and asymmetries in protoplanetary disks are considered as signposts of ongoing planet formation. In this work, we conduct 
three-dimensional radiative transfer simulations to model the intriguing disk around HD\,143006 that has three dust rings and a bright arc. 
A complex geometric configuration, with a misaligned inner disk, is assumed to account for the asymmetric structures. The two-dimensional 
surface density is constructed by iteratively fitting the ALMA data. We find that the dust temperature displays a notable discontinuity at 
the boundary of the misalignment. The ring masses range from 0.6 to $16\,M_{\oplus}$ that are systematically lower than those inferred in 
the younger HL\,Tau disk. The arc occupies nearly 20\% of the total dust mass. Such a high mass fraction of dust grains concentrated in a 
local region is consistent with the mechanism of dust trapping into vortices. Assuming a gas-to-dust mass ratio of 30 that is constant 
throughout the disk, the dense and cold arc is close to the threshold of being gravitationally unstable, with the Toomre 
parameter $Q\,{\sim}\,1.3$. Nevertheless, our estimate of $Q$ relies on the assumption for the unknown gas-to-dust mass ratio. 
Adopting a lower gas-to-dust mass ratio would increase the inferred $Q$ value. Follow-up high resolution observations of dust and gas lines 
are needed to clarify the origin of the substructures.}



\maketitle


\begin{multicols}{2}
\section{Introduction}
 
To date, there are more than 4000 exo-planets discovered using various techniques, such as radial velocity, transit surveys, and 
a recently developed method based on disk kinematics \cite{Mayor1995,Basri2005,Ricker2015,Pinte2019}. How exactly these planets 
form remains one of the major questions in modern astrophysics. One direct way to solve the problem is to image the (proto)planets 
that are still in the making within their natal disks, as they can provide clear insights on where, when, and how planets are born. 
Unfortunately, such observations are very challenging because the embedded planets are much fainter in the optical/infrared regime 
compared to the host central star, and even to the emission knots, if present, from the disk material in the vicinity of the planet. 
Consequently, only a few young stars have been claimed to harbour (proto)planets so far \cite{Keppler2018,Muller2018}, and most of 
them still require further confirmation \cite{Quanz2013,Currie2019,Wagner2019}.  

Nowadays, infrared and millimeter images of disks, taken by the Very Large Telescope Spectro-Polarimetric High-contrast 
Exoplanet REsearch (VLT/SPHERE) and Atacama Large Millimeter/submillimeter Array (ALMA), are able to 
achieve angular resolutions on the order of ${\sim}\,0.05^{\prime\prime}$ \cite{Beuzit2008}, corresponding to spatial scales 
of ${\sim}\,7\,\rm{AU}$ at the distance of nearby star formation regions (${\sim}\,140\,\rm{pc}$, for instance Taurus). 
The accumulating SPHERE and ALMA surveys have revealed a variety of disk substructures, such as rings \cite{Avenhaus2018,Long2018,Andrews2018}, 
gaps \cite{ALMA2015,Boekel2017,Andrews2020}, spirals \cite{Benisty2015,Perez2016,Benisty2017,Huang2018a}, and arcs \cite{vandermarel2013,Dong2018,Cazzoletti2018,Perez2018}. 
The interaction between the disk and embedded planets is one of the proposed explanations for the formation of these features \cite{Kley2012}. 
The properties of substructures together with the disk morphology can be used to constrain the parameters of the embedded planets \cite{Fung2015,Lodato2019}. 
Alternative mechanisms, which can produce substructures without planets, include zonal flows generated by the magnetorotational 
instability turbulence \cite{Bai2014,Cui2021}, fast pebble growth near condensation fronts of volatiles \cite{Zhang2015},
operation of a secular gravitational instability \cite{Takahashi2016}, and the pile-up of drifting pebbles \cite{Drazkowska2016, Jiang2021}. 
In these scenarios, disk substructures may play a role of dust trapping, which on the one hand prevents dust particles from 
rapid inwards drift \cite{Weidenschilling1977}, and on the other hand promotes grain growth and concentration to form planetesimals \cite{Pinilla2015}. 
 
It is of particular interest to investigate the properties and stability of disk substructures, because such efforts 
help to understand how the disk evolves and for what timescales the substructures will impact the process of planet formation. 
Arcs, or their analogs, e.g., crescents and horseshoes, are often interpreted as vortices arising from the Rossby wave 
instability (RWI) at the edge of a gap created by a young planet \cite{vandermarel2013,Zhu2014}. Some numerical simulations 
show that, in weakly turbulent disks, vortex-driven arcs and rings might survive for more than 
${\sim}10^3$ planetary orbits, and even for disk evolutionary timescales (${\sim}\,\rm{Myr}$) \cite{Fu2014,Hammer2021,Jiang2021}. 
However, observational constraints on the strength of turbulence exist for only a few disks \cite{Teague2016,Flaherty2017}.

Rings and arcs in some cases appear extraordinarily far from the central star \cite{Dong2018,Benisty2018}. In such cold 
and dense regions, gravitational instability may start to operate, which in turn complicates the determination for 
the fate of substructures. HD\,143006 stands out from this disk category because of its complex geometry and very 
bright arc outside of the millimeter dust disk \cite{Benisty2018,Perez2018}. This is a G7 young star located 
in the Upper Scorpius association at a distance of 166\,pc \cite{Luhman2012,GAIA2018}. Its effective temperature 
and luminosity are ${\sim}\,5623\,\rm{K}$ and ${\sim}\,3.8\,L_{\odot}$, respectively \cite{Barenfeld2016,Andrews2018}. 
Comparing the location in the Hertzsprung-Russell diagram with models of pre-main-sequence evolutionary tracks returns 
a stellar mass of ${\sim}\,1.77\,M_{\odot}$, and an age of $4\,{-}\,12\,\rm{Myr}$ \cite{Andrews2018}. 
In this work, we focus on this intriguing object, and conduct three-dimensional radiative transfer modeling of 
the spectral energy distribution (SED), and high resolution infrared and millimeter images, with the goal of 
constraining the dust density and temperature distributions. On the basis of the modeling result, we investigate 
the properties and gravitational stability of the substructures.

\section{Observational data}
\label{sec:obs}

\begin{figure}[H]
 \centering
 \includegraphics[scale=0.43]{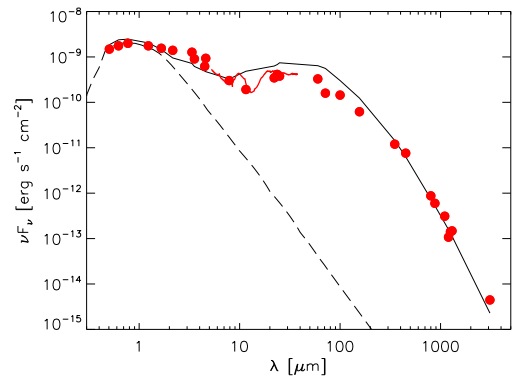}
 \caption{(Color online) SED of the HD\,143006 disk. The red dots refer to photometric data points at different wavelengths, whereas the red solid 
 line shows the Spitzer/IRS spectrum. The black solid line stands for the best-fit model with its parameter set given in \cref{tab:paras}.
 The dashed line represents the input photospheric model from the Kurucz database.} 
 \label{fig:sed}
\end{figure}

The SED has been widely used to constrain the grain size distribution and overall structure of the disk \cite{Robitaille2007}. 
We compiled the photometry of HD\,143006 from the GAIA \cite{GAIA2018}, 2MASS \cite{Cutri2003}, WISE \cite{Cutri2013}, and IRAS 
catalogs \cite{Neugebauer1984}. In addition, individual flux measurements from optical to millimeter regimes reported 
in the literature are also collected \cite{Grankin2007,Carpenter2008,Sylvester1996,Natta2004,Barenfeld2016,Andrews2018}. 
As shown in \cref{fig:sed}, the SED shows a clear dip at $\lambda\,{\sim}\,10\,\mu{\rm m}$, suggesting that substructures 
(e.g., gaps) exist in the distribution of micron-sized dust grains. A linear fit to the data points longer than $0.8\,\rm{mm}$ results 
in a millimeter spectral index of $\alpha\,{\sim}\,2.9$, which yields a dust opacity slope of 0.9 under an assumption 
of optically thin emission in the Rayleigh-Jeans limit \cite{Miyake1993,Andrews2007}. This value is lower than that of 
the interstellar medium dust (i.e., ${\sim}\,1.7$), indicating that dust grains in the disk
have already grown to millimeter sizes \cite{Ricci2010,Testi2014}. 

\begin{figure}[H]
 \centering
 \includegraphics[scale=0.43]{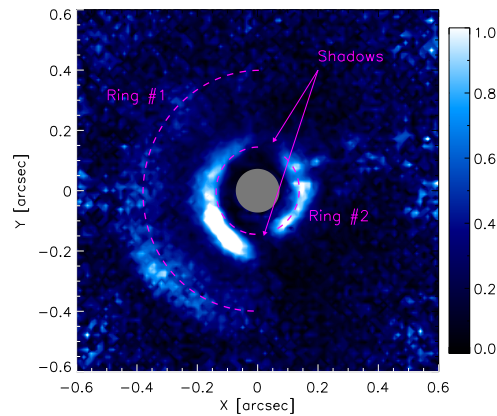}
 \caption{(Color online) J-band polarized intensity from SPHERE observations. The quantity shown here is actually the intensity 
  scaled by $r^2$, where $r$ is the distance to the central star. The region masked by the coronagraph (with a diameter of 145 mas) is 
  indicated by the gray circle. The magenta dashed curves depict the position and extent of two rings. The arrows point to two 
  shadows that are located at $\rm{PA}\,{\sim}\,190^{\circ}$ and $340^{\circ}$. The nomenclature of these substructures is directly 
  taken from Benisty et al. \cite{Benisty2018}.} 
 \label{fig:sphere}
\end{figure}

\begin{figure*}[t]
 \centering
 \includegraphics[scale=0.6]{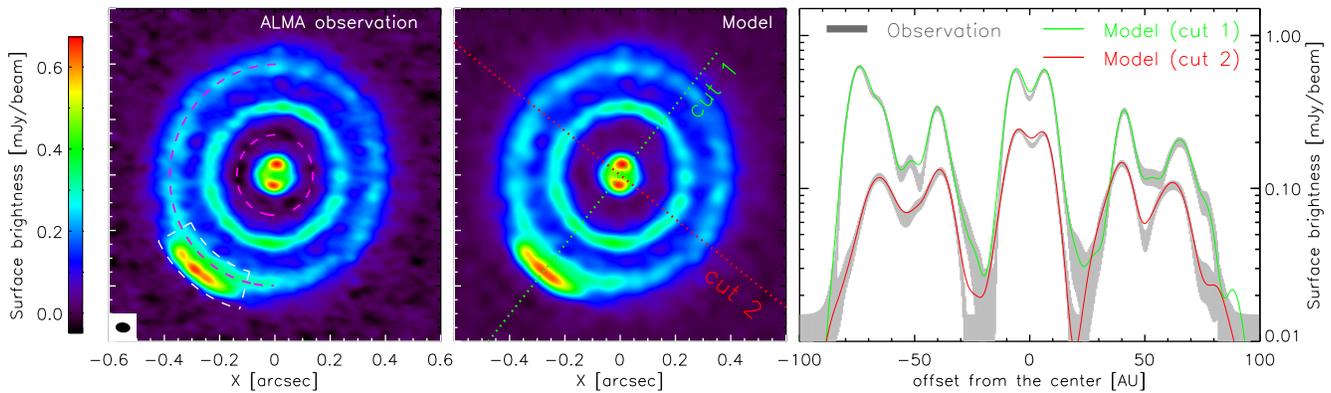}
 \caption{(Color online) {\it Left panel:} fiducial image of the Band 6 ALMA observation generated by the DSHARP team \cite{Andrews2018}. 
 The region enclosed with white dashed lines defines the arc structure in this work. The boundaries are described as $59.7\,{\leq}\,R\,{\leq}\,83\,\rm{AU}$ 
 and $118^{\circ}\,{\leq}\,{\rm PA}\,{\leq}\,165^{\circ}$. The magenta dashed curves depict the position and extent of two rings identified from the 
 SPHERE observation, see \cref{fig:sphere}. {\it Middle panel:} best-fit model image. The green and red dotted lines 
 depict two cut directions ($\rm{PA}\,{=}\,141.5^{\circ}$ and $51.5^{\circ}$), along which the observational and model surface brightnesses 
 are extracted and compared in the {\it right panel}. The thickness of the grey lines represents the uncertainties of the observation. 
 Both the model and observed surface brightness of the cut 2 are scaled down by a factor of two to have a better visualization.} 
 \label{fig:mmimage}
\end{figure*}

Using the VLT/SPHERE instrument, Benisty et al. \cite{Benisty2018} observed HD\,143006 at an angular resolution of ${\sim}\,0.037^{\prime\prime}$. 
\cref{fig:sphere} shows the J-band ($\lambda_{0}\,{\sim}\,1.25\,\mu{\rm m}$) polarized intensity, which traces the scattered 
light from micron-sized dust grains populating the disk's surface layer \cite{Benisty2018}. The disk reveals 
a strong east/west brightness asymmetries, and two rings. The outer Ring {\#}1 extends from $R\,{\sim}\,0.3^{\prime\prime}$ 
to $0.5^{\prime\prime}$, and appears obscure in the western side. The inner Ring {\#}2 has a central position 
of ${\sim}\,0.15^{\prime\prime}$. Along the azimuthal direction, its brightness is not continuously distributed, but shows 
two shadows at $\rm{PA}\,{\sim}\,190^{\circ}$ and $340^{\circ}$. These prominent features are indicative of a complex disk 
configuration with a misaligned inner disk \cite{Min2017}. The clear signature of flux depletion between rings (i.e., in the gap)
and also in the central cavity is consistent with the dip shown in the mid-infrared SED. 

HD\,143006 was selected as one of the 20 disks observed in the Disk Substructures at High Angular Resolution Project (DSHARP) \cite{Andrews2018}. 
With an angular resolution of ${\sim}\,0.05^{\prime\prime}$, the 1.25\,mm continuum image obtained with ALMA is resolved into three 
rings (B8, B40, and B64) at radial locations of $R\,{\sim}\,0.05^{\prime\prime}$, $0.24^{\prime\prime}$ and $0.39^{\prime\prime}$
respectively, see the left panel of \cref{fig:mmimage}. Note that the nomenclature of these rings is directly taken 
from P\'erez et al. \cite{Perez2018}. Unlike the J-band scattered light, millimeter dust rings are broadly symmetric,
except that the innermost B8 ring features two peaks along a PA of ${\sim}\,165^{\circ}$.  Ring {\#}2 identified in the SPHERE image lies 
closer to the central star than the B40 ring, whereas the position of the SPHERE Ring {\#}1 matches well with the B64 ring. 
The most distinct feature in the ALMA image is the bright arc in the southeast, located just exterior to the B64 ring, and 
extends from ${\sim}\,118^{\circ}$ to ${\sim}\,165^{\circ}$ in the azimuthal direction. This arc has a counterpart in the SPHERE 
image, but it shows a broader radial and azimuthal extent. 

\section{Radiative transfer modeling}
\label{sec:rtmodel}

The origin of the rich substructures of HD\,143006 is yet not fully understood. Hydrodynamic simulations demonstrate that 
an inclined equal mass binary causes disc breaking into two distinct annuli, with the resulting inner disk misaligned with 
the outer region \cite{Facchini2018,Benisty2018}. This scenario can produce azimuthal asymmetries shown in the scattered light.
Ballabio et al. \cite{Ballabio2021} found that besides an inclined binary, a planetary companion located 
at $R\,{=}\,32\,\rm{AU}$ is required to simultaneously explain the morphological features in the scattered light and 
the millimeter dust rings. We do not attempt to re-investigate the same story. Instead, we are interested to constrain 
the temperature and density distribution of the disk with detailed radiative transfer modeling, and to infer the 
properties of the substructures. 

\subsection{Disk configuration} 

\begin{figure}[H]
 \centering
 \includegraphics[scale=0.4]{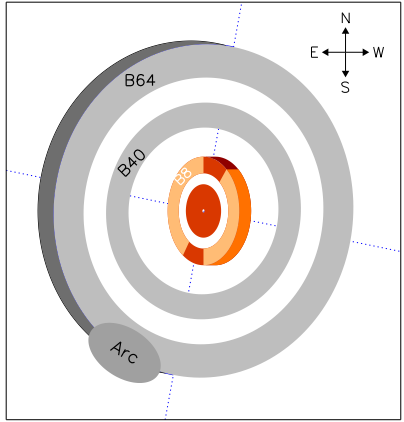}
 \caption{(Color online) None-scaled cartoon of the disk configuration. The inner (red color) and outer (grey color) disks are misaligned with 
 an angle of $\varpi$, see \cref{tab:paras}. The east part of the outer disk is the near side to us, whereas the near side of the inner 
 disk is in the west. The blue dotted lines mark the major and minor axis of the outer disk. Prominent substructures discussed 
 in \cref{sec:obs} are labeled out. The B8 ring exhibits two peaks along a PA of ${\sim}\,165^{\circ}$, which are indicated with 
 the two dark red regions.} 
 \label{fig:sketch}
\end{figure}

\cref{fig:sketch} shows the configuration of our model. The inner and outer disks do not share the same geometry. They 
are misaligned with an angle of $\varpi\,{=}\,30^{\circ}$, which can be roughly constrained by the location 
of the shadows \cite{Min2017}. The near side of the outer disk is in the east, whereas the western part of the inner disk 
is closer to us. The inclination and position angle of the outer disk are $18.6^{\circ}$ and $169^{\circ}$, respectively \cite{Huang2018b}, 
while for the inner disk we adopt an inclination of $11.4^{\circ}$, and PA of $0^{\circ}$. The border between the inner 
and outer disks is fixed to $R_{\rm b}\,{=}\,18\,\rm{AU}$, which is not well constrained. \cref{tab:paras} summarizes 
the values of these geometric parameters that are all kept constant in the fitting process for simplicity. 

\subsection{Dust density distribution} 

To account for the complex geometry of the HD\,143006 disk, we introduce three-dimensional radiative transfer models that include 
two distinct dust grain populations, i.e., a small grain population (SGP) and a large grain population (LGP). The model is established 
under the spherical coordinates. The resolution in the radial ($R$), poloidal ($\theta$), and azimuthal directions ($\phi$) 
are 256, 1024, and 181 cells, respectively. Grid refinement is considered in the vicinity of the border ($R_{\rm b}$) between 
the inner and outer disks as well as the location of the rings. This is done to ensure a smooth transition from optically thin to 
thick cells, meaning that these disk regions are well resolved by the grid.

The inner and outer radii, and total dust mass of the disk are denoted as $R_{\rm in}$, $R_{\rm out}$ 
and $M_{\rm dust}$, respectively. The SGP occupies $f_{\rm SGP}\,{=}\,15\%$ of the total dust mass, and 
its scale height follows a power-law profile:
\begin{equation}
h(R) = h_{100}\times\left(\frac{R}{100\,\rm{AU}}\right)^\beta.
\label{eqn:heightgas}
\end{equation}
The parameter $\beta$ quantifies the disk flaring, and $h_{100}$ stands for the scale height at $R\,{=}\,100\,\rm{AU}$. 
Conversely, the LGP dominates the dust mass, i.e., $0.85M_{\rm dust}$, and is concentrated close to 
the midplane with a scale height of $\Lambda\,h$, where the settling parameter $\Lambda\,{=}\,0.2$. Such a vertical 
stratification of grain size distributions has been commonly used to mimic the effect of dust settling 
in protoplanetary disks \cite{Dullemond2004,Dalessio2006,Liu2019}. Observationally, it is difficult to constrain the 
total mass of small grains characterized by $f_{\rm SGP}$, because infrared observations only probe a portion of small grains 
in the surface layer of the disk. The parameter $\Lambda$ is highly related to the strength of turbulence that is poorly constrained
from observations \cite{Teague2016}. In our models, $h_{100}$, $\beta$ and $\Lambda$ work together to define the scale height of 
millimeter dust grains that basically determines the characteristic temperature of millimeter continuum emission. As $h_{100}$ and
$\beta$ are treated as free parameters (see \cref{sec:betarange}), the effect of a mild change to $\Lambda$ on the millimeter continuum 
can be roughly compensated by according variations in $h_{100}$ and/or $\beta$ that result in a similar millimeter dust scale height. 
Therefore, to reduce the total number of free parameters, $\Lambda$ was fixed during the fitting procedure. We directly took the values 
for $f_{\rm SGP}$ and $\Lambda$ that were adopted in the literature modeling of protoplanetary disks \cite{Andrews2011,Fedele2018,Liu2019}.

The volume density of the dust grains is parameterized as 
\begin{equation}
\rho_{\rm{SGP}}(R,z,\phi)\,{=}\,\frac{f_{\rm SGP}\,{\times}\,\Sigma_{\rm dust}(R,\phi)}{\sqrt{2\pi}\,h}\,\exp\left[-\frac{1}{2}\left(\frac{z}{h}\right)^2\right], \\
\label{eqn:sgp}
\end{equation}
\begin{equation}
\rho_{\rm{LGP}}(R,z,\phi)\,{=}\,\frac{(1-f_{\rm SGP})\,{\times}\,\Sigma_{\rm dust}(R,\phi)}{\sqrt{2\pi}\,{\times}\,\Lambda\,h}\,\exp\left[-\frac{1}{2}\left(\frac{z}{\Lambda\,h}\right)^2\right], \\
\label{eqn:lgp}
\end{equation}
where $\Sigma_{\rm dust}(R,\phi)$ is the dust surface density profile, and varies both in the radial and azimuthal axis.
Previous studies on axisymmetric disks usually parameterize $\Sigma_{\rm dust}$ with an analytic formula, for example 
a power law or a power law plus an exponential taper \cite{Lynden-Bell1974,Andrews2011}. However, in radiative transfer 
models, linking such smooth (or clean) surface densities to high resolution ALMA images is impractical 
because the wealth of fine-scale features and asymmetries are extremely hard to be captured. Instead, to construct the 
surface density profile particularly for HD\,143006, we develop an iterative fitting loop (see \cref{sec:surdens}), which 
is improved upon the algorithm that has been successfully applied to a few other targets \cite{Liu2019}.  

According to \cref{eqn:sgp,eqn:lgp}, SGP and LGP follow a same surface density pattern. In principle, the distributions of 
small and large grains are different, because large grains gradually decouple from the gas as the disk evolves, while small grains 
are expected to be well-mixed with the gas. Generally, one can scale down the gas distribution to represent the surface density profile of SGP. 
However, constraining the gas surface density requires high resolution and high sensitivity observations 
of multiple gas lines \cite{Isella2016}, which are currently not available for HD\,143006. Moreover, although the quality of the SPHERE 
data is high, translating polarized scattered light map to surface densities is not straightforward, because the morphology and intensity of infrared 
polarization are highly dependent on the polarization efficiency, e.g., grain composition, shape and size distribution. Unfortunately, until now, we 
have poor knowledge on these dust properties in protoplanetary disks. Since our work focuses more on the ALMA data that probe large dust grains, 
for simplicity, we assumed that the surface densities of small and large grains follow the same profile. Such an assumption is not expected 
to have a significant impact on the properties (e.g., mass, millimeter optical depth, and Toomre Q parameter) of the disk substructures 
that will be derived and discussed in \cref{sec:resdis}.

\begin{table}[H]
\footnotesize
\begin{threeparttable}
\caption{Overview of the model parameters.}
\label{tab:paras}
\doublerulesep 0.1pt \tabcolsep 5pt 
\begin{tabular}{lccc}
\toprule
 Parameter      & Explored values    &  Best fit         &  Comment     \\
 \hline
 $T_{\rm eff}$\,[K]          & 5623  &  5623             &   fixed  \\
 $L_{\star}\,[L_{\odot}]$    & 3.8   &  3.8              &   fixed  \\
 $D$\,[pc]                   & 166   &  166              &   fixed  \\
 $A_{\rm V}$\,[mag]          & 0.4   &  0.4              &   fixed  \\
 $R_{\rm in}$\,[AU]          & 0.2   &  0.2              &   fixed  \\
 $R_{\rm out}$\,[AU]         & 100   &  100              &   fixed  \\
 $R_{\rm b}$\,[AU]           & 18    &  18               &   fixed  \\
 $\Lambda$                   & 0.2   &  0.2              &   fixed  \\
 $f_{\rm SGP}$               & 0.15  &  0.15             &   fixed  \\
 $\Sigma_{\rm dust}\,[\rm{g/cm^2}]$     &  see \cref{sec:surdens}   &  see \cref{fig:surdens}  &   free   \\
 $\beta$              & 1.05, 1.1, 1.15, 1.2, 1.25 &  $1.1_{-0.05}^{+0.1}$  &   free  \\
 $H_{100}$\,[AU]      & 4, 6, 8, 10, 12       &  $10_{-2}^{+0}$           &   free  \\
 $M_{\rm dust}\,[10^{-5}\,M_{\odot}]$ & $-$   &  $9.7_{-0.3}^{+1.3}$ \tnote{(a)} & $-$   \\
 $a_{\rm max.SGP}$\,[$\mu{\rm m}$] &  2  &    2              &  fixed \\
 $a_{\rm max.LGP}$\,[mm]           &  3  &    3              &  fixed \\     
 $i_{\rm inner}\,[^{\circ}]$     & 11.4  &   11.4           &  fixed \\
 ${\rm PA_{inner}\,[^{\circ}]}$  & 0     &   0              &  fixed \\
 $i_{\rm outer}\,[^{\circ}]$     & 18.6  &   18.6           &  fixed \\
 ${\rm PA_{outer}\,[^{\circ}]}$  & 169   &   169            &  fixed \\
 $\varpi\,[^{\circ}]$            & 30    &   30             &  fixed \\
\bottomrule
\end{tabular}
\begin{tablenotes}
\item[(a)] The total dust mass $M_{\rm dust}$ is obtained by integrating the surface density $\Sigma_{\rm dust}$ that 
is constructed in the fitting procedure. Hence, $M_{\rm dust}$ is not a direct fitting parameter.
\end{tablenotes}
\end{threeparttable}
\end{table}

\subsection{Dust properties}
\label{sec:dustmodel}
Birnstiel et al. \cite{Birnstiel2018} proposed a dust model to interpret the ALMA data of the DSHARP disks. We directly 
take their model that consists of water ice, astronomical silicates, troilite, and refractory organic material, with 
volume fractions being 36\%, 17\%, 3\% and 44\%, respectively. The bulk density of the dust mixture 
is $\rho_{\rm grain}={\rm 1.675\,g\,cm^{-3}}$. We use the Mie theory to calculate the mass absorption/scattering coefficients. 
The grain size distribution follows the power law ${\rm d}n(a)\propto{a^{-3.5}} {\rm d}a$ with a 
minimum grain size fixed to $a_{\rm{min}}=0.01\,\mu{\rm m}$. The maximum grain size is set to $a_{\rm{max}}\,{=}\,2\,\mu\rm{m}$ 
for the SGP. To account for the effect of grain growth implied by the shallow millimeter spectral index (see \cref{sec:obs}), we 
adopt $a_{\rm{max}}\,{=}\,3\,\rm{mm}$ for the LGP.

Zhu et al. \cite{Zhu2019} showed that taking dust scattering into account is important to estimate the total amount of 
dust mass in protoplanetary disks. Full scattering matrices are included in our radiative transfer models to treat 
dust scattering in a more realistic way.

\subsection{Stellar heating}
\label{sec:heating}
For simplicity, only stellar irradiation is considered as the heating source of the disk. The stellar parameters are listed 
in \cref{tab:paras}. The photospheric spectrum is taken from the Kurucz database \cite{Kurucz1994}, assuming ${\rm log}\,g\,{=}\,3.5$ 
and solar metallicity. We found that a mild extinction of $A_{\rm V}\,{=}\,0.4\,\rm{mag}$ is needed to match the 
optical part of the SED. The extinction law is taken from Cardelli et al. \cite{Cardelli1989}.

\subsection{Fitting procedure}

\subsubsection{Explored ranges for $\beta$ and $H_{100}$}
\label{sec:betarange}

There are three free parameters/quantity in the model: $\beta$, $H_{100}$, and $\Sigma_{\rm dust}$. 
The disk flaring $\beta$ and scale height $H_{100}$ define the vertical structure of the disk. The location of 
shadows seen in the SPHERE observation is sensitive to the height of the scattering surface $z_{\rm scat}$ as
a function of $R$. Using the equation developed in Min et al. \cite{Min2017}, Benisty et al. \cite{Benisty2018} 
estimated $z_{\rm scat}/R\,{=}\,0.12$ at $R_{\rm b}\,{=}\,18\,\rm{AU}$. Assuming that the scattering surface is 
higher than the scale height $h$ by a factor of ${\sim}\,2–4$, the disk aspect ratio $h/R\,{\sim}\,0.03\,{-}\,0.06$. 
We note that such a calculation is highly dependent on the adopted geometric parameters and dust temperature 
as well, and therefore the result can only be treated as a guide. 

In the fitting process, we explored a grid of values for $\beta$ and $H_{100}$ that are within the ranges of 
[1.05, 1.25] and [4\,AU, 12\,AU], respectively. Using \cref{eqn:heightgas}, the deduced aspect ratio $h/R$ 
spans a relatively broader range of 0.03$\,-\,$0.1. 

\subsubsection{Building the two-dimensional surface density}
\label{sec:surdens}
The most difficult part of the modeling is to construct the surface density profile $\Sigma_{\rm dust}(R,\phi)$. This 
is due to three reasons. First, it is naturally understood that for HD\,143006, $\Sigma_{\rm dust}$ varies not 
only in the radial direction, but also with azimuthal angle. Second, what the ALMA image (see the 
left panel of \cref{fig:mmimage}) shows is the millimeter continuum emission from a three dimensional disk 
projected onto the plane of sky. The projection parameters (i.e., inclination and position angle) are different 
between the inner and outer disks. Third, there is a misalignment complicating the geometry further.   

To deal with the difficulty, we devised an iterative fitting loop. An initial surface density profile $\Sigma_{\rm ini}(R,\phi)$ should be given. 
In practice, it can be an arbitrary form, for instance a power law. The surface density is closely linked to the ALMA observation 
due to the fact that the optical depth at millimeter wavelengths is low. Therefore, we extract the surface brightness on the ALMA image 
along 181 azimuthal angles ($\phi$) distributed from 0 to $360^{\circ}$. The observed two-dimensional surface brightness 
distribution ${\rm SB}_{\rm obs}(R,\phi)$ was first deprojected using the geometric parameters given in \cref{tab:paras}, and then 
imported into the radiative transfer model as the initial surface density $\Sigma_{\rm ini}(R,\phi)$. 
After that, the iteration starts, and is followed by the steps itemized below.

\begin{itemize}
\item[(a)] We perform the radiative transfer simulation using the well-tested code \texttt{RADMC-3D}\footnote{http://www.ita.uni-heidelberg.de/~dullemond/software/radmc-3d/.} \cite{radmc3d2012}. The raytraced model image at 1.25\,mm is obtained.
\item[(b)] To simulate the ALMA observation, we compute the model visibilities at the same uv coordinates of the observation. 
           Using the Common Astronomy Software Applications (CASA) package \cite{McMullin2007}, the resulting model visibilities was CLEANed 
		   with the same parameters used for HD\,143006 in the DSHARP project \footnote{https://almascience.org/alma-data/lp/DSHARP}. 
\item[(c)] On the CLEANed image, we extracted the surface brightness ${\rm SB}_{\rm mod}(R,\phi)$ at the same 
           locations as what we did on the observed ALMA map.
\item[(d)] A ratio $\xi(R,\phi)$ as a function of radius and azimuthal angle is calculated through dividing the observed brightness 
           by the model brightness: $\xi(R,\phi)\,{=}\,{\rm SB}_{\rm obs}(R,\phi)/{\rm SB}_{\rm mod}(R,\phi)$. We deprojected $\xi(R,\phi)$
		   using the disk geometric parameters to get a scaling factor $\xi^{\rm D}(R,\phi)$ for the next iteration.
\item[(e)] The surface density profile $\Sigma_{\rm dust}(R,\phi)$ used as the input for the model is scaled by the point-by-point ratio $\xi^{\rm D}(R,\phi)$.
           The fitting loop goes back to step (a).
\end{itemize}
		      
The iteration continues	until the changes in the model surface brightness are less than 5\% at all locations. This is typically 
satisfied after ${\sim}\,20$ steps. Once the convergence is achieved, we also simulated the SED, and polarized intensity map 
at $1.25\,\mu{\rm m}$, in addition to the millimeter image. 

\begin{figure*}[t]
 \centering
 \includegraphics[scale=0.55]{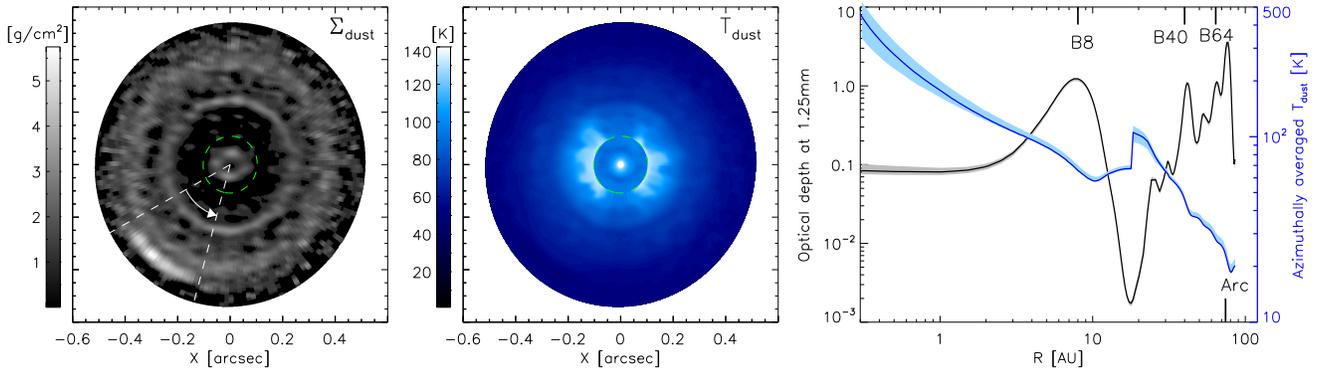}
 \caption{(Color online) {\it Left panel:} two-dimensional surface density profile of the best-fit model. {\it Middle panel:} two-dimensional temperature distribution
 of the best-fit model. The green dashed circles mark the border ($R_{\rm b}\,{=}\,18\,\rm{AU}$) between the inner and outer disk. {\it Right panel:} optical depth 
 at 1.25\,mm ($\tau_{\rm 1.25\,mm}$, black line) and mass-averaged mean dust temperature ($T_{\rm dust}$, blue line) as a function of 
 radius. To highlight the properties towards the arc, we actually show an azimuthal average on $\tau_{\rm 1.25\,mm}$ and $T_{\rm dust}$ within 
 $\rm{PA}\,{=}\,118^{\circ}$ and $165^{\circ}$, see the white curved arrow in the left panel.}
 \label{fig:surdens}
\end{figure*}

\subsubsection{Evaluating the quality of fit}
\label{sec:fitquality}
It can be seen from \cref{tab:paras} that there are 25 different combinations of $\{\beta,H_{100}\}$. For each of 
these possibilities, we separately ran the above-mentioned iteration from scratch until they converge. In the end, 
we obtained 25 models, and their final converged surface densities $\Sigma_{\rm dust}(R,\phi)$ (therefore, the total 
dust mass in the disk $M_{\rm dust}$) are different from each other.

Near-infrared scattered light is produced by small dust grains in the surface layer of the disk. We first compare the 
morphological features in the polarized intensity map between the model and SPHERE observation. \cref{fig:irmodimg} shows 
four models as an illustration. The effect of shadowing cast by the misalignment is very sensitive to the aspect ratio.
We found that models with aspect ratios within $0.07\,{\lesssim}\,h/R\,{\lesssim}\,0.10$ are able to reproduce the overall 
morphology of the SPHERE image. \cref{tab:fitres} summarizes the fitting results. There are 9 models with $h/R$ in the satisfied 
range. 

Then, we evaluate the quality of fit to the SED and ALMA data for the 9 models. The goodness of fit is given by 
\begin{equation}
\chi^2\,{=}\,\sum_{i=1}^{N}\left(\frac{\mu_{i}-\omega_{i}}{\sigma_{i}}\right).
\end{equation}
For the SED, $N$ is the number of observed wavelengths, $\omega_{i}$ are the
photometric measurements, and $\mu_{i}$ are the predicted flux densities at
these wavelengths. The observational flux uncertainties are denoted as $\sigma_{i}$. 
For the ALMA image, $N$ is the number of pixels taken into account, $\omega_{i}$ are the
observed surface brightness in each pixel, and $\mu_{i}$ are the model surface brightness. 
The quantity $\sigma_{i}$ now is the rms noise of the ALMA image, which we fixed to 
$0.015\,\rm{mJy/beam}$ for every pixel \cite{Andrews2018}. The total goodness of fit 
is evaluated via 
\begin{equation}
\chi^2_{\rm total}\,{=}\,g\cdot\chi^2_{\rm SED}+(1-g)\cdot\chi^2_{\rm ALMA},
\end{equation}
where $g$ is a weighting factor that is adjustable to make a compromise for the quality of fit between the SED and 
ALMA data. There is not an universal value for $g$. The $\chi^2_{\rm SED}$ and $\chi^2_{\rm ALMA}$ values for each model 
are listed in \cref{tab:fitres}. As can be seen, $\chi^2_{\rm SED}$ is roughly an order of magnitude lower 
than $\chi^2_{\rm ALMA}$. To balance the weighting between these two independent assessments, $g$ was taken to be 0.9. 
However, we emphasize that our choice is only applicable for this particular study. \cref{fig:sedmodels} shows how 
the SED varies with the geometric parameters of the disk. The best fit is identified as the one with the minimum 
$\chi^2_{\rm total}$. The uncertainties of the best-fit parameter set are deduced from the models 
with $\chi^2_{\rm total}$ lower than 1.1 times the minimum $\chi^2_{\rm total}$.

\section{Results and discussion}
\label{sec:resdis}
The fitting results are presented in \cref{fig:sed,fig:mmimage,fig:irmodimg}, whereas \cref{tab:paras} 
summarizes the corresponding parameter set. Our model well matches the ALMA data and the 
morphology of the infrared polarized intensity map, see for instance how high the quality 
of fit is for the arc and rings in the right panel of \cref{fig:mmimage}.   

The model SED is broadly consistent with the observation, except for the data points 
at ${\sim}\,3\,\mu{\rm m}$ and ${\sim}\,40\,\mu{\rm m}$. The geometry of the inner disk 
for HD\,143006 might be more complicated than what we assumed (see \cref{fig:sketch}), 
for instance there might be an puffed-up inner rim \cite{Dullemond2001}. 
Moreover, a variability in accretion or an inclined companion can lead to changes 
in the structure of the inner disk \cite{Flaherty2012}. These factors can explain the 
discrepancies in the near-infrared flux. For simplicity, we assume that the flaring 
index $\beta$ and scale height $H_{100}$ are the same in the inner and outer disk. 
The mild overestimation of far-infrared flux can be reconciled by reducing $H_{100}$ (or $\beta$,
or both) for the outer disk while maintaining the vertical structure of the inner disk. 
Nevertheless, such finely refinements are not expected to have a significant impact on 
the overall properties (i.e., temperature, surface density, and Toomre parameter) of 
the disk.   

\begin{figure*}[t]
 \centering
 \includegraphics[scale=0.6]{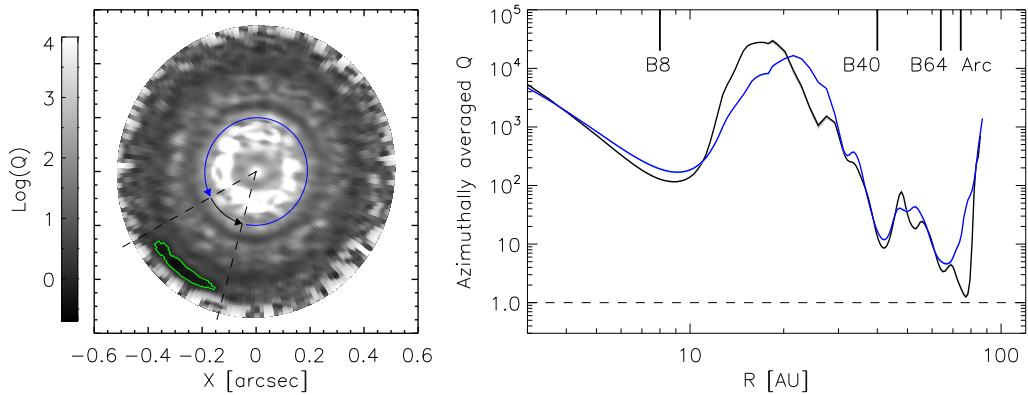}
 \caption{(Color online) {\it Left panel:} two-dimensional distribution of the Toomre parameter $Q$, assuming a constant gas-to-dust mass ratio of 30 throughout the disk. 
 The green curve indicates the contour level of $Q\,{=}\,1$. The black and blue curved arrows mark two azimuthal ranges (i.e., $118\,{\leq}\,{\rm PA}\,{\leq}\,165^{\circ}$ and $-195\,{\leq}\,{\rm PA}\,{\leq}\,118^{\circ}$, respectively), within which an average on $Q$ is calculated separately, and then compared in the {\it right panel}.} 
 \label{fig:toomre}
\end{figure*}

The left panel of \cref{fig:surdens} shows the surface densities projected onto the plane of sky. 
A projection is performed to have a one-to-one comparison with the observation. The reconstructed surface densities show 
obvious fluctuations along both the radial and azimuthal directions. We checked the quality of the ALMA 
data, and found that in most disk regions (excepting for the gap around $R\,{\sim}\,22\,\rm{AU}$), the
signal-to-noise ratio is higher than 3. In the arc and rings where we are most interested, the ratio is higher than ${\sim}\,10$.  
These mean that the derived surface densities for the arc and rings are reliable since our iterative fitting is driven by the data. 
Nevertheless, surface density fluctuations in the gap around $R\,{\sim}\,22\,\rm{AU}$ may reflect the noise of the ALMA observation.
The right panel presents the optical depth at 1.25\,mm and the mass-averaged temperature, derived from an azimuthal average 
between ${\rm PA}\,{=}\,118^{\circ}$ and $165^{\circ}$. As can be seen, all the three ALMA rings (B8, B40, B64) are marginally 
optically thick: $\tau_{1.25\,\rm{mm}}\,{\sim}\,1$. This is different from the result presented by 
Huang et al. \cite{Huang2018b}, who obtained an optical depth of $\tau_{1.25\,\rm{mm}}\,{\lesssim}\,0.2$. 
When calculating $\tau_{1.25\,\rm{mm}}$, they assumed a temperature profile being the simplified expression for a passively 
heated, flared disk: $T\,{\propto}\,R^{-0.5}$ \cite{Chiang1997,Dullemond2001}. 
It is clear that such an approximation is not sufficient for HD\,143006, see the blue solid 
line in the figure. Instead, due to the misalignment, the temperature exhibits a discontinuity 
at the boundary ($R_{\rm b}\,{=}\,18\,\rm{AU}$) between the inner and outer disk. Along the radial  
direction towards the arc (${\rm PA}\,{=}\,141.5^{\circ}$), stellar photons can directly impinge upon 
the inner rim of the outer disk, resulting in a sharp increase in temperature. Beyond that 
radius, the temperature decreases by roughly following a power law with $T\,{\sim}\,20\,\rm{K}$ 
at the arc location. Moreover, in the analysis of Huang et al. \cite{Huang2018b}, dust scattering 
is not taken into account. The difference in optical depth between our work and theirs can be understood 
in terms of these two reasons. The bright arc is optically thick with $\tau_{1.25\,\rm{mm}}\,{\sim}\,2.7$. 

\begin{table}[H]
\footnotesize
\begin{threeparttable}
\caption{Dust mass and mean temperature in the substructures}
\label{tab:massres}
\doublerulesep 0.1pt \tabcolsep 8pt 
\begin{tabular}{lccc}
\toprule
 Substructure & Region & Dust mass  &  Mean $T_{\rm dust}$     \\
              & [AU or $^{\circ}$]  & [$M_{\oplus}$] & [K]     \\
 \hline
 B8  & $4.6\,{\leq}\,R\,{\leq}\,10.7$  &  $0.6_{-0.1}^{+0.0}$  & $74.1_{-6.1}^{+0.5}$  \\
 B40 & $35\,{\leq}\,R\,{\leq}\,44.8$   &  $3.9_{-0.0}^{+0.4}$  & $48.0_{-3.0}^{+0.0}$  \\
 B64 & $52.5\,{\leq}\,R\,{\leq}\,74.6$ &  $15.5_{-0.3}^{+1.8}$ & $28.7_{-2.4}^{+0.6}$  \\
 \multirow{2}{*}{Arc} & $59.7\,{\leq}\,R\,{\leq}\,83$  &  \multirow{2}{*}{$6.3_{-0.2}^{+0.9}$} & \multirow{2}{*}{$23.1_{-2.0}^{+0.5}$} \\
                      & $118\,{\leq}\,{\rm PA}\,{\leq}\,165$ &  &  \\
\bottomrule
\end{tabular}
\end{threeparttable}
\end{table}

The misalignment leads to a highly asymmetric temperature distribution, see the middle panel of \cref{fig:surdens}. 
The eastern/western sides of the outer disk's rim (located at $R_{\rm b}\,{=}\,18\,\rm{AU}$, the green dashed circle in the plot) 
are directly exposed to the central star. The typical temperature there reaches $110\,\rm{K}$. Conversely, the 
northern/southern parts of the rim are cool with $T_{\rm dust}\,{\sim}\,60\,\rm{K}$. \cref{tab:massres} gives the 
mean temperature of each substructure. We defined the arc region as the area enclosed with the 
white dashed lines in the left panel of \cref{fig:mmimage}. The radial range of the arc overlaps with 
that of the B64 ring, because we considered that both features are tightly connected. Adopting the 
condensation temperature reported in the literature \cite{MartinDomenech2014,Luna2014,Zhang2015}, the B8, B40, 
and B64 rings are associated with the ice lines of ${\rm NH_3}$, ${\rm H_2S}$, and ${\rm CH_4}$, respectively. By integrating the surface 
density, we obtained the dust mass in the substructures. The total dust mass of the best model 
is $M_{\rm dust}\,{=}\,9.7\times10^{-5}\,M_{\odot}$ (or $32\,M_{\oplus}$). The arc contains 
about 20\% of the dust mass. This number should be regarded as a lower limit 
due to the optical thickness $\tau_{1.25\,\rm{mm}}\,{\sim}\,2.7$. The mass fraction is lower 
than that of the arc in HD\,135344B (i.e., ${\sim}\,50\%$), which shares a similar disk morphology \cite{Cazzoletti2018}.
The ring masses range from 0.6 to $16\,M_{\oplus}$. They are systematically lower 
than those derived in the younger (${\sim}\,1\,\rm{Myr}$) HL\,Tau disk. In HL\,Tau, the dust masses 
in the rings are from 16 to ${\sim}\,120\,M_{\oplus}$ \cite{Liu2017}. If the rings are created 
by planets, this may suggest that much of the dusty material in the HD\,143006 disk have already been 
converted to the mass of planets (or planetary cores).   

As can be seen from \cref{tab:massres}, the rings, particularly the B64 ring, and the arc occupy a large amount of dust mass. 
In addition, the arc is located so distant from the central star that the temperature is as low as $23\,\rm{K}$. Gravitational 
instability may be in action at these cold and dense regions. We checked this possibility using the Toomre criterion \cite{Toomre1964}.
The Toomre $Q$ parameter is defined as
\begin{equation}
Q=\frac{c_{s}\Omega_{k}}{\pi G \Sigma_{\rm gas}},
\end{equation}
where $c_s$ is the local sound speed, $\Omega_{K}(R)$ is the angular frequency, and $\Sigma_{\rm gas}$ stands for the gas surface density. 
In terms of $Q$, a disk/substructure is unstable to its own self-gravity if $Q\,{<}\,1$, and stable if $Q\,{>}\,1$. As there is no 
constraints on $\Sigma_{\rm gas}$, for simplicity, we scaled up the constructed $\Sigma_{\rm dust}$ by a constant 
gas-to-dust mass ratio of 30 that is found to be representative via millimeter dust and gas observations of a sample of protoplanetary 
disks \cite{Williams2014,Ansdell2016,Ansdell2017}. The two-dimensional distribution of $Q$ is shown in the left panel of \cref{fig:toomre}. 
The value is below unity at the center of the arc, as indicated by the green contour. The right panel of \cref{fig:toomre} displays the azimuthally 
averaged $Q$ as a function of radius. Two ranges of azimuthal angles were considered in order to highlight the difference 
at various locations. While most regions including the three rings are gravitationally stable, $Q$ approaches ${\sim}1.3$ at 
the radial position of the arc. Pohl et al. \cite{Pohl2015} showed that an embedded massive planet (planet-to-star mass 
ratio of 0.01) can induce gravitational collapses in massive disks when $Q$ is around unity. Nevertheless, analyzing 
infrared interferometric data excludes the presence of a companion with mass ratios of ${\gtrsim}\,0.1$ \cite{Benisty2018}. 
Future searches for less massive (proto)planets with high-contrast imaging are desired to test the hypothesis.

The gas-to-dust mass ratio reflects how quickly the dust and gas components dissipate, in particular relative to each other.  
The inherited interstellar medium gas-to-dust mass ratio is 100. ALMA observations of dust and gas lines reveal that 
the ratio can drop to 10 in some disks \cite{Ansdell2016,Soon2019}. Arcs are often interpreted as a result of dust trapping into 
vortices that preferentially concentrate dust grains. Numerical simulations of vortex formation and dust trapping have shown that the 
gas-to-dust mass ratio in the center of vortices can decrease down to 10 and even ${\sim}\,1$ \cite{Meheut2012,Sierra2017}. 
In such a situation, $Q$ will be above unity, and gravitational instability is suppressed accordingly. Future high resolution 
and high sensitivity observations of gas lines are required to constrain the gas surface density, and therefore the 
gas-to-dust mass ratio, particularly in the local arc region. Furthermore, hydrodynamic simulations dedicated on 
HD\,143006, with the inclusion of the feedback of dust onto gas \cite{Crnkovic2015}, will help to understand the 
fate of the arc.       

Vortices are regions acting as local pressure maxima in the disk, and they can form at the edge of a planet-induced gap 
due to the excitation of RWI. An important prediction of dust trapping toward vortices is that larger dust grains are expected 
to be concentrated into a narrower region \cite{Birnstiel2013,Lyra2013}, which has been observationally identified in several 
disks, e.g., Oph IRS 48 \cite{vandermarel2015}, HD\,142527 \cite{Casassus2015}, MWC\,758 \cite{Marino2015}, and 
HD135344B \cite{Cazzoletti2018}. Investigating the spatial distribution of dust grains with different sizes requires high 
resolution observations at multi-wavelengths. Due to the lack of such datasets, we distributed the SGP and LGP 
into the same azimuthal extent in the radiative transfer models. Future high resolution observations at different wavelengths, 
and follow-up search for (proto)planets are indispensable to clarify the nature of the arc in the HD\,143006 disk.

\section{Summary}

We performed three dimensional radiative transfer modeling of the HD\,143006 disk that shows three dust rings and a bright arc
in the outermost region on the DSHARP ALMA map. The goal of our work is to constrain the density and temperature distributions, 
and to investigate the possibility of triggering a gravitational instability in the substructures. In order to account for the 
ALMA and SPHERE observations, our model features a complex geometry with a misalignment between the inner and outer disk.

All of the three rings (B8, B40, B64) are marginally optically thick ($\tau_{1.25\,\rm{mm}}\,{\sim}\,1$) at the observed ALMA wavelength, 
while the optical thickness of the arc is $\tau_{1.25\,\rm{mm}}\,{\sim}\,2.7$. Our self-consistent modeling gives higher optical depth 
than those derived from a simple analytic method in the literature. We found that the temperature as a function of radius resembles  
two connecting power laws with a sharp jump at the boundary of the misalignment. The temperature in the rings are from 74 to 
29\,K, and it drops down to 23\,K in the arc. The total dust mass in the disk is about $32\,M_{\oplus}$, with the arc 
sharing about 20\% of the amount. The dust masses in the rings range from $0.6\,M_{\oplus}$ to $16\,M_{\oplus}$, which are systematically 
lower than those found in the younger HL\,Tau disk. 

While all the rings are gravitationally stable, the Toomre parameter $Q$ in the arc is around unity when assuming a gas-to-dust mass ratio 
of 30 that is constant in the entire disk. The arc is close to the threshold of gravitational collapse in this case. If gravitational instability is 
in action and the cooling is efficient enough, giant planets may quickly form \cite{Mayer2002}. However, if the arc indeed traces a vortex, 
millimeter dust grains preferentially concentrate in the arc, which will decrease the local gas-to-dust mass ratio. Given that the gas-to-dust mass 
ratio is an unknown assumption, adopting a lower value would increase our estimate of $Q$. In this situation, gravitational instability will 
not take place. Future gas and dust observations at high resolutions are required to better constrain the gas-to-dust mass ratio, and to 
clarify the origin of the arc in the HD\,143006 disk.    

\Acknowledgements{We thank the anonymous referees for their constructive comments that highly improved the manuscript. 
We sincerely thank Myriam Benisty for sharing the reduced SPHERE data of HD\,143006. Y.L. acknowledges the financial support 
by the Natural Science Foundation of China (Grant No. 11973090). M.F. acknowledge funding from the European Research Council (ERC) under the 
European Union's Horizon 2020 research and innovation program (grant agreement No. 757957). We acknowledge the science research grants from 
the China Manned Space Project with NO. CMS-CSST-2021-B06. {\it Herschel} is an ESA space observatory with science instruments provided by 
European-led Principal Investigator consortia and with important participation from NASA. ALMA is a partnership of ESO (representing its 
member states), NSF (USA) and NINS (Japan), together with NRC (Canada), MOST and ASIAA (Taiwan), and KASI (Republic of Korea), in cooperation 
with the Republic of Chile. The Joint ALMA Observatory is operated by ESO, AUI/NRAO and NAOJ. This work has made use of data from the 
European Space Agency (ESA) mission Gaia (https://www.cosmos.esa.int/gaia), processed by the Gaia Data Processing and Analysis Consortium (DPAC, https://www.cosmos.esa.int/web/gaia/dpac/consortium). Funding for the DPAC has been provided by national institutions, in particular the 
institutions participating in the Gaia Multilateral Agreement.}

\InterestConflict{The authors declare that they have no conflict of interest.}



\bibliographystyle{scichina}
\bibliography{hd143006}

\begin{appendix}

\renewcommand{\thesection}{Appendix}

\section{More information about the fitting results}
\label{sec:irimage}

\begin{figure*}[t]
\centering
\includegraphics[scale=0.5]{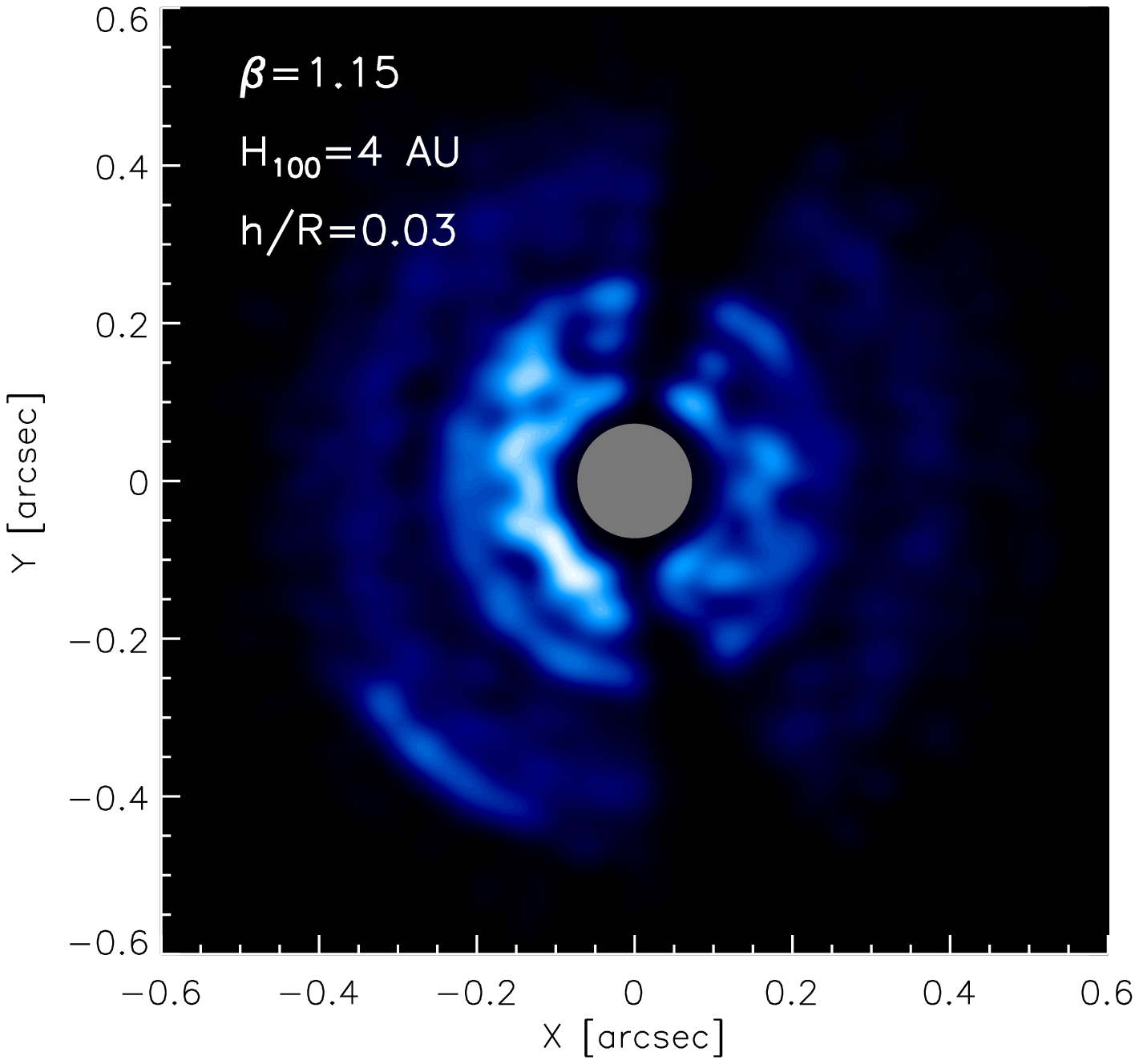}
\includegraphics[scale=0.5]{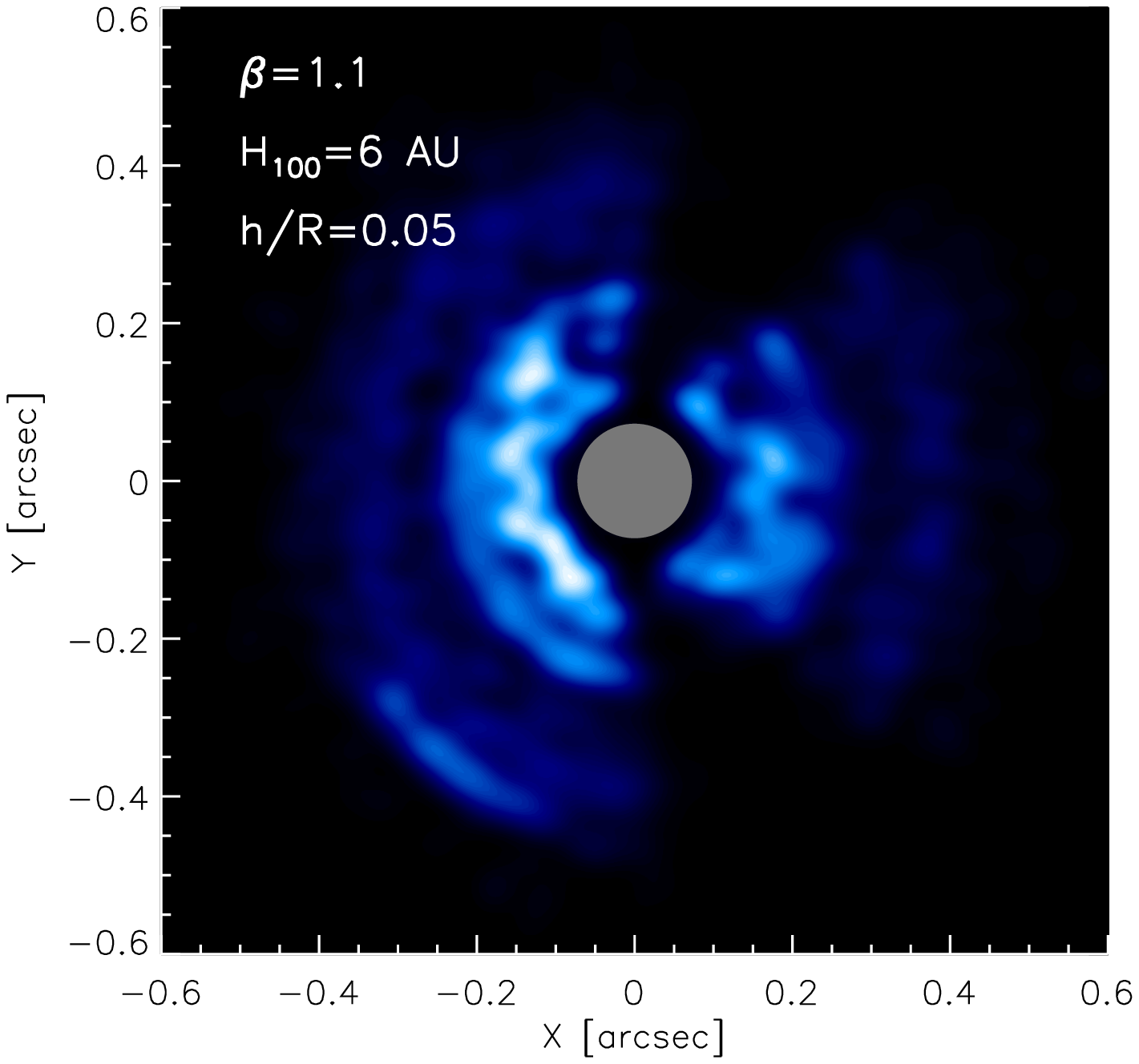}
\includegraphics[scale=0.5]{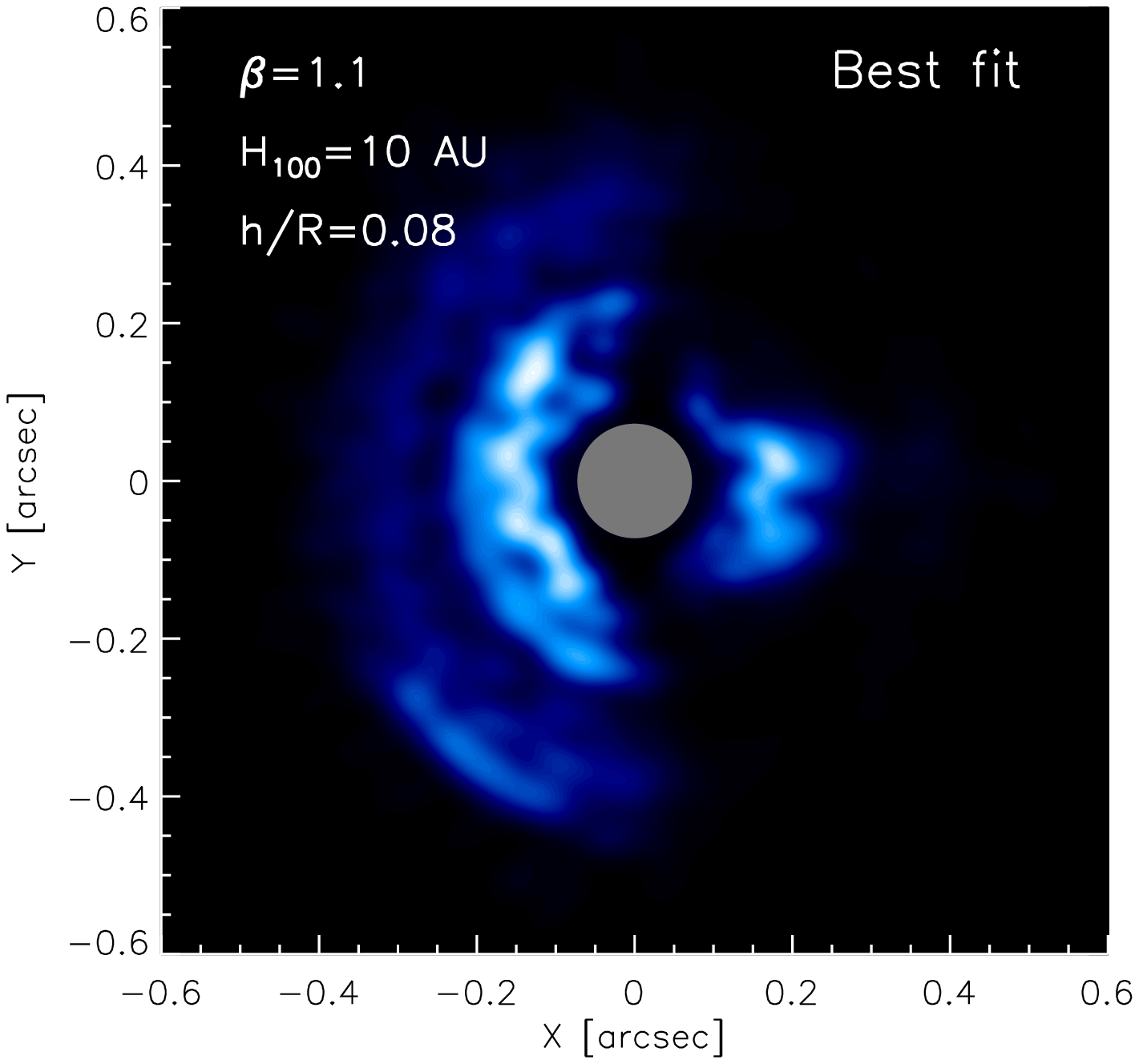}
\includegraphics[scale=0.5]{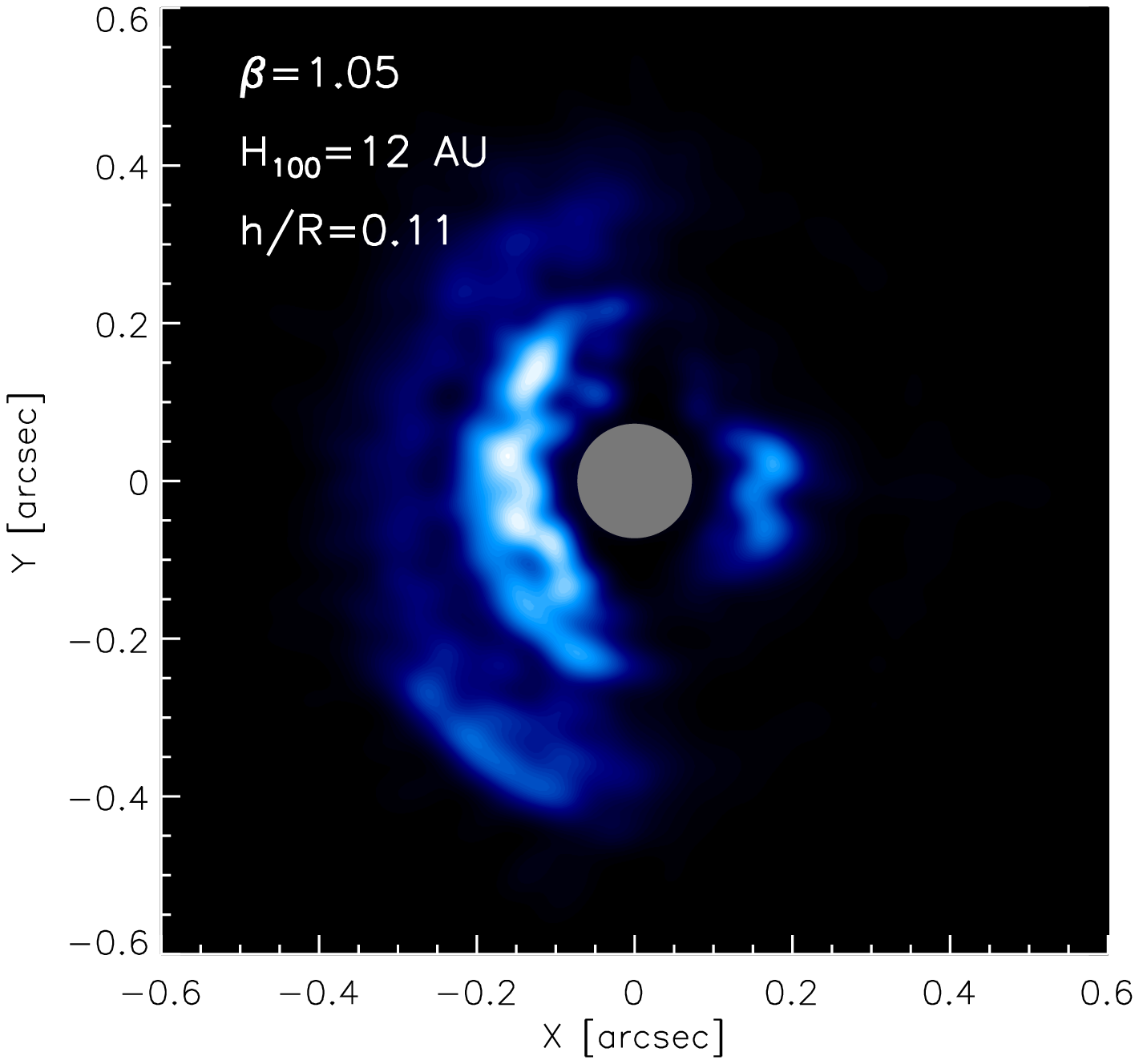}
\caption{(Color online) model images of the polarized intensity at $1.2\,\mu{\rm m}$ for different aspect ratios of the disk. The intensity is scaled by $r^2$ to 
compensate for the drop off in stellar flux. The corresponding SEDs of these models are shown in \cref{fig:sedmodels}.} 
\label{fig:irmodimg} 
\end{figure*}

The location of shadows and the overall morphology of the scattered light map are dependent on the geometric parameters 
of the disk. In the radiative transfer analysis, we fixed the inclination and position angle of the inner and outer 
disks, and the misalignment angle as well. Hence, the remaining freedom characterizing the morphology of the polarized 
intensity map is the aspect ratio that is determined by $\beta$ and $H_{100}$ via \cref{eqn:heightgas}.
\cref{fig:irmodimg} shows how the morphological features change with the aspect ratio $h/R$. 
We found that disk configurations with aspect ratios $0.07\,{\lesssim}\,h/R\,{\lesssim}\,0.10$ are able 
to interpret the overall morphology of the SPHERE image. There are 9 models with $h/R$ being in this range.
\cref{tab:fitres} summarizes the fitting results for each model. The millimeter optical depth, dust temperature, 
and the Toomre $Q$ parameter of the arc are also provided.

\begin{figure*}[t]
\centering
\includegraphics[scale=0.5]{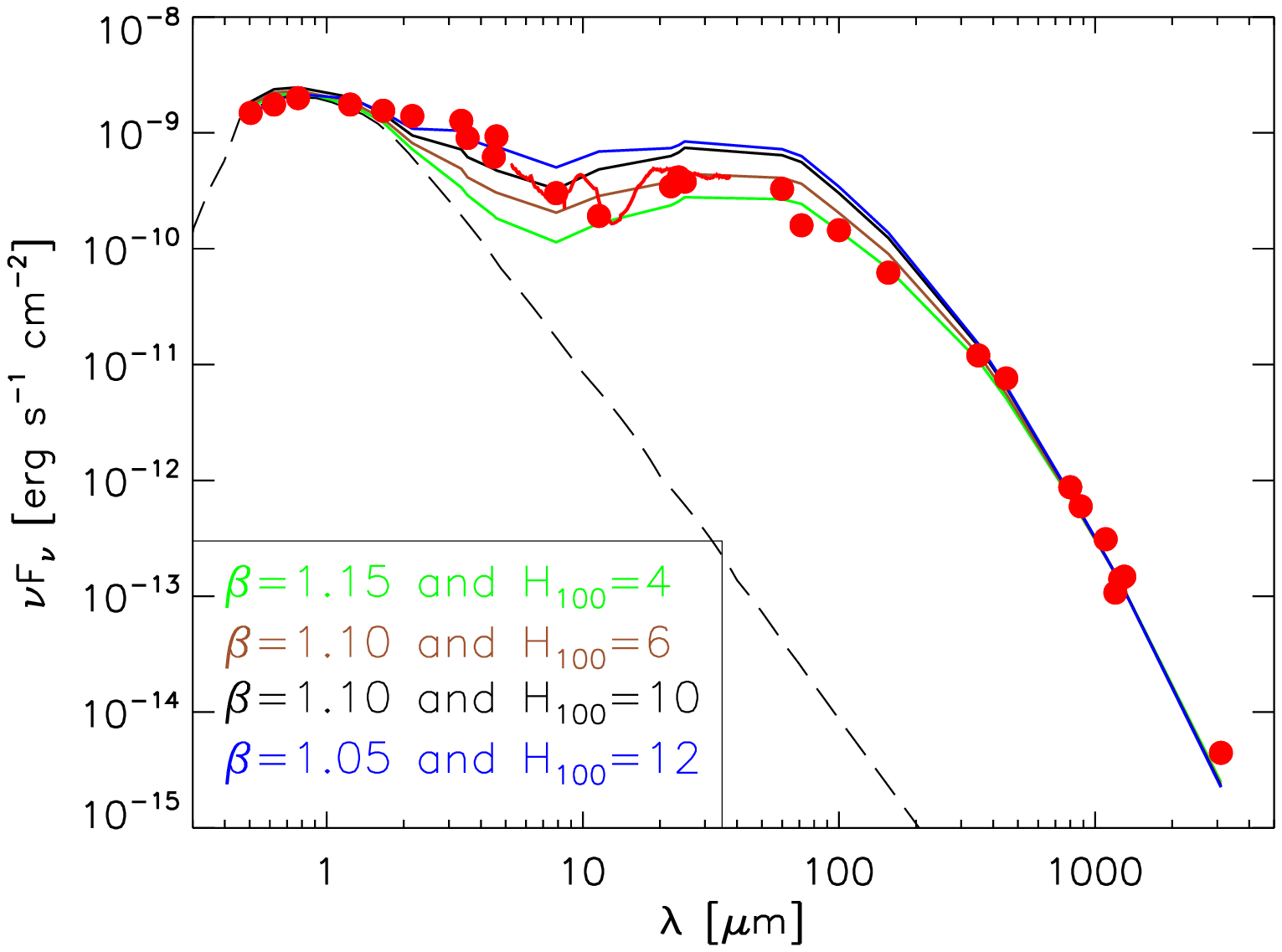}
\caption{(Color online) SEDs of models with different $\beta$ and $H_{100}$. The corresponding scattered light images of these models can be found in \cref{fig:irmodimg}.} 
\label{fig:sedmodels} 
\end{figure*}

The parameters $\beta$ and $H_{100}$ work together to determine the vertical structure of the disk, and therefore 
affect the level of infrared excess. Generally, a larger $\beta$ and/or $H_{100}$ produces more infrared excess, and vice versa. 
The four models shown in \cref{fig:sedmodels} illustrate this tendency, and they have the same parameter values  
to the ones presented in \cref{fig:irmodimg}.

\newpage

\begin{table*}[t]
\footnotesize
\begin{threeparttable}
\caption{Fitting results of each model.}
\label{tab:fitres}
\doublerulesep 0.1pt \tabcolsep 5pt 
\begin{tabular}{cccccccccccc}
\toprule
 Model ID    &  $\beta$  &  $H_{100}$  &  $h/R$  &  $M_{\rm dust}$  &  Match with SPHERE data$^{\,(a)}$  & $\chi^2_{\rm SED}$  &  $\chi^2_{\rm ALMA}$  &  $\chi^2_{\rm total}$  &  $\tau_{\rm 1.25mm}$(arc)  &  
 $T_{\rm dust}$(arc)     &  $Q$(arc)  \\
             &    & [AU]  &  & $[10^{-5}\,M_{\odot}]$  &  &   &  &  &  &   [K]   &       \\
 
\hline
 01          & 1.05   &  4     &  0.037      &  15.0 &  N  & $-$ & $-$   &   $-$  &  $-$  &  $-$ &  $-$ \\
 02          & 1.1    &  4     &  0.034      &  14.0 &  N  & $-$ & $-$   &   $-$  &  $-$  &  $-$ &  $-$ \\
 03          & 1.15   &  4     &  0.031      &  14.0 &  N  & $-$ & $-$   &   $-$  &  $-$  &  $-$ &  $-$ \\
 04          & 1.2    &  4     &  0.028      &  13.0 &  N & $-$  & $-$   &   $-$  &  $-$  &  $-$ &  $-$ \\
 05          & 1.25   &  4     &  0.026      &  13.0 &  N & $-$  & $-$   &   $-$  &  $-$  &  $-$ &  $-$ \\
 06          & 1.05   &  6     &  0.055      &  12.0 &  N & $-$  & $-$   &   $-$  &  $-$  &  $-$ &  $-$ \\
 07          & 1.1    &  6     &  0.051      &  12.0 &  N & $-$  & $-$   &   $-$  &  $-$  &  $-$ &  $-$ \\
 08          & 1.15   &  6     &  0.046      &  12.0 &  N & $-$  & $-$   &   $-$  &  $-$  &  $-$ &  $-$ \\
 09          & 1.2    &  6     &  0.043      &  11.0 &  N & $-$  & $-$   &   $-$  &  $-$  &  $-$ &  $-$ \\
 10          & 1.25   &  6     &  0.039      &  11.0 &  N & $-$  & $-$   &   $-$  &  $-$  &  $-$ &  $-$ \\
 11          & 1.05   &  8     &  0.073      &  11.0 &  Y & 2878 & 33692 & 5959 & 3.1 & 21.1 & 1.1 \\
 12          & 1.1    &  8     &  0.067      &  11.0 &  N & $-$  & $-$   &   $-$  &  $-$  &  $-$ &  $-$ \\
 13          & 1.15   &  8     &  0.062      &  10.0 &  N & $-$  & $-$   &   $-$  &  $-$  &  $-$ &  $-$ \\
 14          & 1.2    &  8     &  0.057      &  10.0 &  N & $-$  & $-$   &   $-$  &  $-$  &  $-$ &  $-$ \\
 15          & 1.25   &  8     &  0.052      &  10.0 &  N & $-$  & $-$   &   $-$  &  $-$  &  $-$ &  $-$ \\
 16          & 1.05   &  10    &  0.092      &  9.8  &  Y & 2909 & 35634 & 6181 & 2.7 & 22.9 & 1.2 \\
 17          & 1.1    &  10    &  0.084      &  9.7  &  Y & 2528 & 35822 & 5857 & 2.7 & 23.1 & 1.3 \\
 18          & 1.15   &  10    &  0.077      &  9.6  &  Y & 2771 & 34379 & 5932 & 2.6 & 23.4 & 1.3 \\
 19          & 1.2    &  10    &  0.071      &  9.4  &  Y & 3095 & 34292 & 6215 & 2.6 & 23.6 & 1.3 \\
 20          & 1.25   &  10    &  0.065      &  9.3  &  N & $-$  & $-$   &   $-$  &  $-$  &  $-$ &  $-$ \\
 21          & 1.05   &  12    &  0.11       &  9.1  &  N & $-$  & $-$   &   $-$  &  $-$  &  $-$ &  $-$ \\
 22          & 1.1    &  12    &  0.10       &  9.0  &  Y & 3768 & 38118 & 7203 & 2.4 & 25.1 & 1.3 \\
 23          & 1.15   &  12    &  0.093      &  8.9  &  Y & 3555 & 42960 & 7495 & 2.4 & 25.3 & 1.3 \\
 24          & 1.2    &  12    &  0.085      &  8.8  &  Y & 3762 & 35776 & 6964 & 2.3 & 25.3 & 1.4 \\
 25          & 1.25   &  12    &  0.078      &  8.7  &  Y & 4202 & 32385 & 7020 & 2.3 & 25.5 & 1.4 \\
\bottomrule
\end{tabular}
\begin{tablenotes}
\item[(a)] This column gives the result for a comparison of scattered light images between the model and SPHERE data. If the model fails to reproduce the overall morphological features, ``N'' is assigned. On the contrary, ``Y'' means that the model and observation are broadly consistent with each other, see \cref{sec:fitquality}.
\end{tablenotes}
\end{threeparttable}
\end{table*}

\end{appendix}

\end{multicols}
\end{document}